\documentstyle[12pt]{article}

\begin{document}

\renewcommand{\theequation}{\arabic{equation}}

\title{The annealed positions of ferromagnetic bonds\\
doped into a 2D antiferromagnet}

\author{N. M. Salem and R. J. Gooding \\
{\em Department of Physics, Queen's University,}\\ 
{\em Kingston, Ontario K7L 3N6} }

\date{\today}

\maketitle

%\newpage

\begin{abstract}
We determine where a given concentration of ferromagnetic (FM) bonds
doped into a square lattice antiferromagnet must go to minimize the 
system's total magnetic energy.  We find (i) an infinite 
degeneracy of ground--state arrangements of FM bonds that correspond 
to completely unfrustrated configurations for classical spins, and 
(ii) this degeneracy is lifted when quantum fluctuations are 
included, and phase separated ground states, such as periodic
arrays of stripes of FM bonds, are found.  
A discussion of the application of these ideas to doped cuprate
high $T_c$ superconductors with annealed disorder is presented.

\end{abstract}
\newpage

% ** Introduction **
% ----------------
%\section{Introduction:}
%\label{sec:intro}

Carriers in the high T$_c$ compounds can be introduced into the 
ubiquitous ${\rm CuO_2}$ planes of the insulating 
antiferromagnetically (AFM) ordered parent compounds either by doping 
with immobile dopants, or by altering the oxygen concentration. 
If, for example, one is concerned with the interpretation of 
experimental studies of the Bednorz--M\"uller high T$_c$ compound 
${\rm La_{2-x}Sr_xCuO_{4+\delta}}$, the carriers in the planes will 
be affected by the {\em quenched} disorder that is present as a result 
of the immobile ${\rm Sr}$ dopant ions.  In contrast to this situation, 
excess oxygen ions ($\delta > 0$) in the 
${\rm x = 0}$ material are mobile at sufficiently high temperatures, 
and thus the disorder introduced by them during the crystal growth 
process can be {\em annealed}, effectively restoring translational 
periodicity of a perfect crystal \cite {excessO}.
The absence of dopant disorder leads to a simplified system, one that 
should be amenable to comprehensive theories of the holes in the 
${\rm CuO_2}$ planes.

In this report we concern ourselves with the possibility of charge
inhomogeneities of the holes in the ${\rm CuO_2}$ planes in the
absence of dopant disorder. Such
structures are potentially related to the frustrated phase separation
phenomenology of Emery and Kivelson \cite {emery} in the form
of striped regions rich in the holes, and many theoretical studies
consistent with such charge distributions have been found via 
Hartree--Fock studies \cite {HF}. However, the mechanism associated
with this microsegregation of holes remains unclear, and thus
in this report we eliminate the kinetic energy of the holes and consider
localized oxygen holes, thus focussing entirely on the
magnetic interactions present in the doped system. 
Due to the annealing away of disorder that is
possible for the super--oxygenated compounds, we imagine that
the localized holes can find the annealed positions that they would assume to
minimize the doped plane's energy. Very similar physics is found in
superoxygenated ${\rm La_2NiO_{4+\delta}}$, although the holes
are believed to exist as partially localized Zhang--Rice singlets and not
as completely localized oxygen holes \cite {lanio}.
For our simplified model we explain how magnetic interactions alone
can produce striped ground states --- whether or not this is an important
driving force associated with such charge inhomogeneities remains an 
open question.

The magnetic interactions in a system with completely localized oxygen
holes is as follows: A localized oxygen hole neighbouring a ${\rm Cu~3d^9}$ 
ion leads to a Kondo--like coupling between the ${\rm S = {1\over 2}}$ Cu
ion and the ${\rm S= {1\over 2}}$ oxygen ion.
Regardless of the sign of this interaction, the coupling
of the oxygen ion to both ${\rm Cu}$ sites leads to
a net ferromagnetic (FM) interaction between the pair of 
${\rm Cu}$ sites surrounding the oxygen hole, an idea
first put forward by Aharony, {\em et al.} \cite {aharony}. Since 
the undoped system corresponds to neighbouring pairs of 
${\rm Cu}$ ions that interact antiferromagnetically, the simplest model
of this system is just that of the so--called Aharony model,
{\em viz.} a distribution of ferromagnetic bonds of
a fixed concentration doped into a square lattice ${\rm S = {1\over 2}}$ AFM.
The Hamiltonian representing this situation is
\begin{equation}
H = J \sum_{<ij>}~{\bf S}_i \cdot {\bf S}_j
- (K + J) {\sum_{<ij>}}^{\prime}~{\bf S}_i \cdot {\bf S}_j
\end{equation}
where $J > 0$ is the AFM exchange constant of the undoped background,
$i$ and $j$ label the lattice sites of an infinite 2D
square lattice, $<ij>$ denotes near neighbours,
$K > 0$ is the ferromagnetic exchange constant for a bond containing
an oxygen hole, and the primed summation indicates that only
those bonds containing an oxygen hole are summed over.
For a random distribution of these FM bonds the background AFM state 
becomes frustrated, leading to a complicated
phase diagram {\em vs.} doping \cite {aharony}.

The fact that we are allowing the oxygen holes to move around and anneal
their positions in the ${\rm CuO_2}$ planes means that the ferromagnetic 
bonds can move in an attempt to lower the system's energy.
Thus, another way of phrasing the question at hand is:
What is the spatial distribution of some fixed concentration of FM
bonds doped into square lattice quantum AFM that minimizes the
system's energy? 
The answer that we have found 
involves the relieving of the frustration induced by the FM
bonds by phase separation. To be specific, for classical
spins a variety of configurations of FM bonds allows for
completely unfrustrated spin textures. Then, when quantum 
fluctuations are included, something we have managed to
do using Colpa's para--diagonalization scheme for the
bilinear, many--boson problem \cite {colpa},
one finds that, {\em e.g.}, for a small FM exchange energy,
stripe configurations are the preferred
low energy state.

Firstly, we describe the solution of this problem for classical
Heisenberg spins (it is to be noted that the following
arrangements of spin directions are also the ground states
for XY and Ising spin systems). Figures 1 $\rightarrow$ 3 show some
of the configurations of FM bonds for which unfrustrated
spin textures may be found. The simplest such configuration is
that shown on the left of Fig. 1 --- we shall refer to this
as the ``plus sign" arrangement. To see that 
an unfrustrated spin texture may be found for this set of
bonds note that with the introduction of
the 4 FM bonds of the plus sign configuration the central
sign is isolated from the AFM background. Thus, by flipping this 
and only this spin, relative to the undoped AFM spin arrangement,
{\em all} FM and AFM bonds in the lattice are satisfied. (One
may generalize the idea of this configuration to isolated core
regions of arbitrary size.)
For a given concentration of FM bonds one may simply place
these plus sign configurations anywhere in the lattice.
Then, for a set ratio of $K/J$, the energies of all such 
arrangements are equal, a fact that
applies to all unfrustrated arrangements that we have found.

There are a variety of other allowed spin textures: The 
(horizontal or vertical) striped phase of width $W$ is shown
on the right of Fig. 1. The unfrustrated spin state is achieved
by flipping all spins on one side of the stripe, beginning
with the sites at the ends of the FM bonds; these
states are unfrustrated for all $W$. Figure 2 shows another
non--frustrating arrangement of FM bonds, a
pair of crossed stripes (here shown for $W = 1$). 
Note that crossed stripes lead to a completely FM region at the
intersection point, and thus may be desirable at large
$K/J$. Figure 3 shows something that we shall refer to as a 
$W = 1$ staircase configuration --- this is somewhat like
a $W = 1$ stripe that is rotated by 45$^\circ$, although
the internal structure of the staircase (a continuous
line of FM bonds) is clearly different from that of
a stripe.

Clearly, there is an infinite degeneracy of such structures
that at $T=0$ minimize the system's energy for classical
spins. However, since all of these structures correspond to
differing local spin environments, when zero--point
spin--wave fluctuations are included the energy of these
configurations for quantum spins will, in general, be different.
We expect that at $T=0$, since these systems are a long way from
any phase transition linear spin--wave theory should
adequately represent the spin excitations.
Then, since spin waves effectively dress a classical state,
we begin with the above determined classical configurations,
on which we then superimpose
spin waves, finally determining which of these
configurations has the 
lowest energy for a given concentration of FM bonds and for a given 
ratio of $K/J$ for quantum spins of length $S$.

It is easy to exclude many configurations from further 
consideration: The inclusion of spin waves means that arrangements 
of FM bonds that are highly localized in space, such as, e.g., the 
plus sign
arrangements, require a large number of spin waves of {\em all}
wave vectors, essentially equally weighted, to represent
the interaction of the surface of the region of FM bonds
with the background AFM environment --- since the short--wavelength
spin waves are of a high energy, they will make configurations that
require their inclusion to be high as well. Thus, we expect and have
limited our theoretical studies to configurations
that are devoid of the plus sign structure. (To confirm the legitimacy
of this approximation we have checked that for all $K/J$ and for the
concentration range of FM bonds that we have studied, periodic
arrangements of plus signs (with square or rectangular unit cells)
for $S = {1\over 2}$ have energies higher than the ground state configurations.) 
Further, in order to access and bias the lowest energy spin waves to
be those dominating the interface regions at which
the FM bonds impinge on the AFM background we consider
{\em periodic arrangements} of the stripes, or crossed stripes,
or staircases.

Once one of these periodic arrangements is selected,
the relationship between the density of FM bonds to the periodic
structure is easily determined. From now on we define the 
concentration
of oxygen holes per ${\rm Cu-O-Cu}$ bond to be $c$. Then, if one
is considering a periodic repetition of stripes of width $W$,
the repeat distance $L$ between stripes must be taken
to be $L = W / (2~c)$. For this particular arrangement this
means that the number of sites in the magnetic unit cell
is $2 \times L$ (or possibly $2 \times 2L$). Thus, standard
Holstein--Primakoff theory of spin waves of the
two--dimensional quantum AFM \cite {stratos} is not directly applicable.
Instead, one must allow for different quantum fluctuations
to exist at every inequivalent site, something
that we have accomplished by using a different boson at
each equivalent site
(where, say, $a_{Ii}^{\dagger}$ represents the creation
of a spin deviation in unit cell $I$ at basis site $i$.)
Due to the large number of sites per unit cell,
the resulting bilinear Hamiltonian involves a very large
number of coupled bosons. We have managed to extract
the spin--wave eigenfrequencies and total system energy for 
this problem using the para--diagonalization formalism developed 
by Colpa --- an extensive discussion of the details of this method 
is contained in Ref. \cite {colpa}.

Our results, for a concentration  $c = 0.1$ and $S = {1\over 2}$, 
are shown in Table I. 
The lowest energy configuration depends on the ratio of $K/J$. 
Since the structure of a system's total energy is 
just $E = -S(S+1)~E_{classical} + S~E_{SWT}$, our results
for the ordering of energies are effectively independent of $S$. Also,
in the concentration range that we have studied, {\em viz.}
$0.083 \leq c \leq 0.125$, there is effectively no dependence of
the ordering of the energies on concentration.

For $K/J$ less than unity, the lowest energy state 
is a $W = 1$ striped phase. 
This is thus an example of magnetic interactions producing
a striped ground state \cite {low}. The driving energy per spin
associated with some kind of stripe ordering is of the order
of 6 K for ${\rm La_2CuO_{4+\delta}}$, and is thus quite small --- this
suggests that interactions such as the Coulomb interaction \cite {low} 
must be included to fully account for the observed charge orderings.
That we find a striped ground state may be understood in
a simple non--interacting approximation: {\em ignoring the interface 
energies} and exactly evaluating the energies of a width $W$ 
stripe with open boundary conditions, and then a width $(L-W)$ 
AFM region, and then weighting the energies of the two regions 
by $W/L$ and
$(L-W)/L$ to finally represent the energy of the bulk lattice,
one finds that the minimum energy for small $K/J$ corresponds to 
$W \rightarrow 0$ (this is simply a statement that for
small $K/J$ FM bonds are not helpful in lowering the system's
energy, no matter where they are put). However, since we are 
imposing that some fixed concentration of FM bonds must be present, 
$W=1$ is the smallest allowed $W$. The $W = 1$ geometry which we find
minimizes the energy for small $K/J$ is that of a periodic repetition
of $W=1$ stripes.

For large $K/J$ we expect an entirely different configuration
to be the minimum energy state, {\em viz.}, if one completely
phase separates the system into a FM region of concentration
$c$, and then a AFM region of concentration $1-c$, {\em and}
is able to produce the unfrustrated interface separating the
two regions, one should find the minimum energy configuration.
A representation of this kind of arrangement is shown in Fig. 4.
Clearly, this is simply the $W \rightarrow \infty$ staircase
with a fixed concentration of FM bonds; alternatively,
this phase is a maximally dense clustering of plus sign structures 
forming a 45$^{\circ}$ interface. 
One may write down an expression for the energy per spin
of this state that is exact in the bulk limit (since the interface 
energy will scale as $1/L$, $N = L^2$ being the number of lattice 
sites) and find
\begin{equation}
{E\over N} = -~2~K~c < {\bf S}_0 \cdot {\bf S}_1 >_{FM}
+~2~J~(1-c) < {\bf S}_0 \cdot {\bf S}_1 >_{AFM}
\end{equation}
Evaluating the near--neighbour correlation functions
for an AFM in the spin--wave approximation
our results for this two--phase 
structure are also listed in Table I.
We see that for $K/J > 1.0$ 
this completely phase separated structure has the lowest 
energy. 

In conclusion, we have made an in depth study of the 
energy--minimizing
configurations of FM bonds doped into a 2D square lattice
AFM. The infinite degeneracy 
of unfrustrated classical spin ground states is lifted 
by zero--point spin--wave excitations which subsequently
leads to striped microsegregated ground states for small $K/J$,
and macroscopically phase separated ground states for
large $K/J$.

\begin{center}
       {\bf  ACKNOWLEDGMENTS}
\end{center}

We wish to thank John Tranquada for many informative conversations
concerning his and Buttrey's experimental work on the nickelate 
system. We also wish to thank Barry Wells, 
Bob Birgeneau and Marc Kastner for helpful comments. 
Much of this work was completed while one of the authors (RJG) was 
visiting the CMSE at MIT, and he wishes to thank them for their warm 
hospitality.  This work was supported by the NSERC of Canada.

\newpage

\title{REFERENCES}

\begin{figure}

\caption{Two frustration--free configurations of FM bonds.
The solid circles represent the transition metal ions
that interact AFM with their nearest
neighbours with strength $J$ in Eq.~(1). The solid
lines are the FM bonds of strength $K$ in Eq.~(1).
On the left of the figure is four FM bonds organized in
the shape of a plus sign. These bonds do not frustrate the AFM
background since the central spin can be flipped over
from the direction it would chose in the undoped AFM 
state.  On the right of the figure is a vertical stripe of FM
bonds, stacked one on top of another. That this
configuration is unfrustrated follows from these changes:
take every spin--up sublattice site on the right of the
stripe and make it a spin--down sublattice. Similarly,
change the down--sublattice sites to be up--sublattice
sites. For an infinite
system, one requires that this stripe be of infinite length
for the total spin texture to be unfrustrated. The
stripes can be of any integral width, $W$, with the figure
displaying a $W = 2$ stripe.}
\end{figure}

\begin{figure}
\caption{The intersection of two $W=1$ stripes,
which also leads to an unfrustrated spin arrangement.
The overlapping portion corresponds to a fully
FM region.}
\end{figure}

\begin{figure}
\caption{A staircase configuration. The solid lines
are seen to create a continuous line of FM bonds,
thus effectively producing a 1D FM chain embedded in
a 2D AFM.}
\end{figure}

\begin{figure}
\caption{A $W \rightarrow \infty$ staircase configuration. This
leads to a two--phase, completely phase separated region
of either FM (lower right) or AFM (upper left) bonds. If the 
interface is taken to be along this direction, the resulting 
ground state spin arrangement is completely unfrustrated.}
\end{figure}

\begin{table}
\caption{The energy per spin, relative to $J$, for $S=1/2$ and $c = 0.1$, 
of some periodic arrangements of classically unfrustrated FM 
configurations for $W=1$, the $W \rightarrow \infty$ crossed stripes
configuration, and the two--phase state shown in Fig. 4. 
The starred energies are the lowest energy configurations for a 
given $K/J$.}

\label{Evsstructure1/2}

%\vskip 0.5 truecm
\begin{center}
\begin{tabular}{cccccc}
\hline \hline
$K/J$ & Striped & Staircase & Crossed Stripes& Crossed Stripes & Staircase\\
  &   $W=1$  & $W=1$  &$W=1$  & $W\rightarrow\infty$ &$W\rightarrow\infty$ \\
\hline
0.1&    -0.6136$^\ast$&    -0.6071& -0.6133 & -0.6095 &   -0.5971~~\\
0.25&   -0.6167$^\ast$&    -0.6117& -0.6163 & -0.6130 &   -0.6046~~\\
0.5&    -0.6243$^\ast$&    -0.6209& -0.6239 & -0.6218 &   -0.6171~~\\
1.0&    -0.6430~~&    -0.6416& -0.6429 & -0.6430 &   -0.6421~~\\
2.0&    -0.6863~~&    -0.6868& -0.6866 & -0.6894 &   -0.6921$^\ast$\\
4.0&    -0.7801~~&    -0.7821& -0.7810 & -0.7864 &   -0.7921$^\ast$\\
10.0&   -1.0740~~&    -1.0763& -1.0754 & -1.0884 &   -1.0921$^\ast$\\
\hline \hline
\end{tabular}
\end{center}
\end{table}


\begin{thebibliography}{sss200}

\bibitem{excessO} D. C. Johnston, {\em et al.},
in {\em Phase Separation in Cuprate Superconductors},
edited by E. Sigmund and K. A. M\"uller, (Springer--Verlag,
Heidelberg, 1994), p. 82;
B. O. Wells, {\em et al.}.  Z. Phys. B {\bf 100}, 535 (1996).

\bibitem{emery} V. Emery and S. Kivelson, High $T_c$ Los Alamos
Proceedings (1994).

\bibitem{HF} See, {\em e.g.}, M. Inui and P. B. Littlewood, Phys.
Rev. B {\bf 41}, 4415 (1991), and references therein.

\bibitem{lanio} J. M. Tranquada, {\em et al.}, Phys. Rev. B {\bf 52},
3581 (1995).

\bibitem{aharony} A. Aharony, {\em et al.},
Phys. Rev. Lett. {\bf 60}, 1330 (1988).

\bibitem{colpa} J. H. P. Colpa, Physica {\bf 93A}, 327 (1978);
{\em ibid}, {\bf 134A}, 377, 1986; 
{\em ibid}, {\bf 134A}, 417, 1986.

\bibitem{stratos} E. Manousakis, Rev. Mod. Phys.
{\bf 63}, 1 (1991).

\bibitem{low} U. L\"ow, {\em et al.}, Phys. Rev. Lett. {\bf 72},
1918 (1994) have found striped ground states in their competing
interactions model only if Coulomb--like interactions are included.

\end{thebibliography}
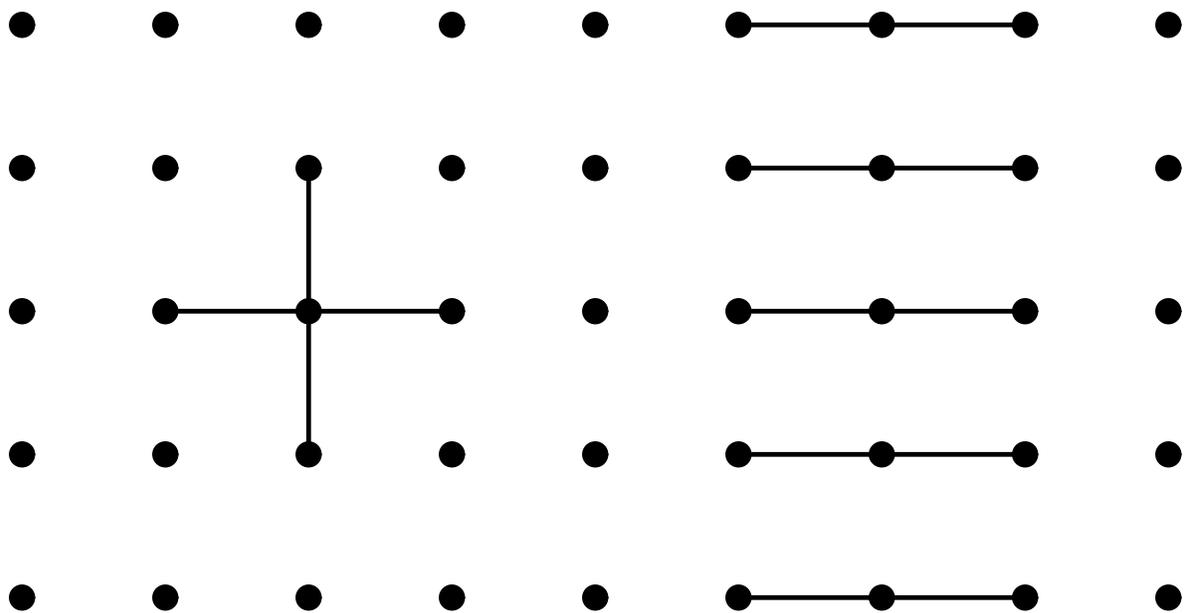
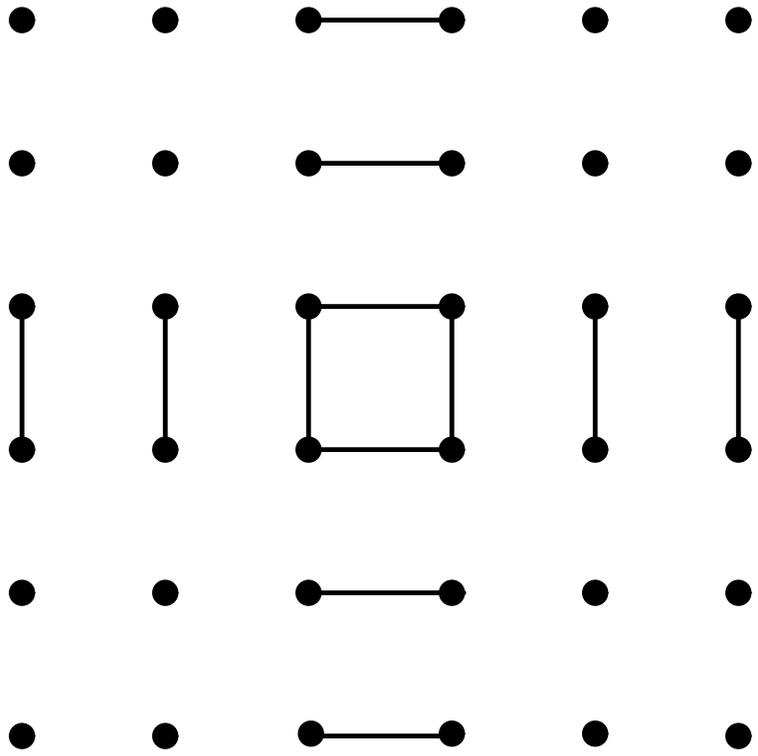
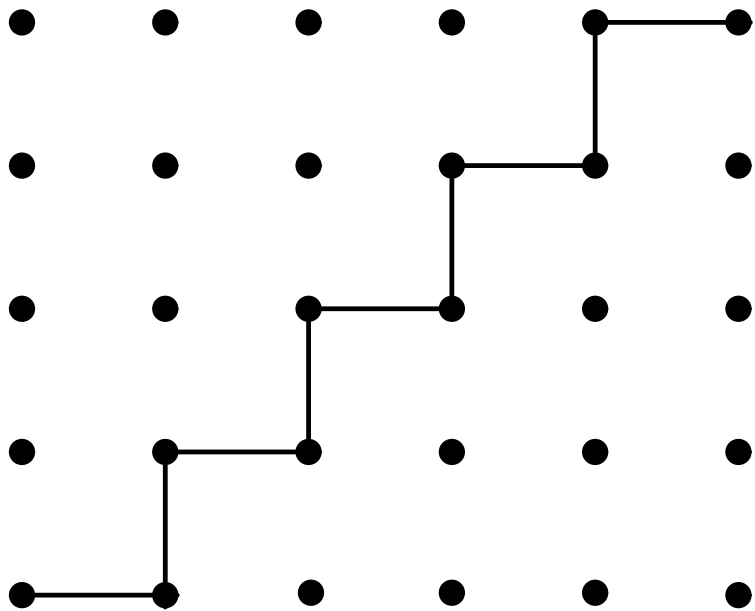
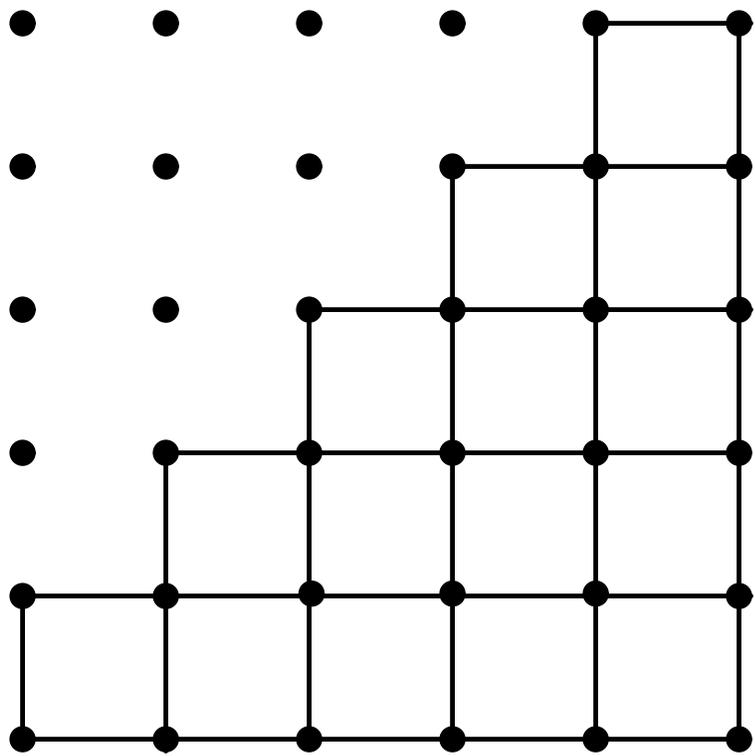
\end{document}